\newcommand\IPA{IPA-CuCl$_3$}
\newcommand\IPACB{IPA-Cu(Cl$_{1-x}$Br$_x$)$_3$}
\newcommand\IPACBS{IPA-Cu(Cl$_{0.95}$Br$_{0.05}$)$_3$}

\documentclass[twocolumn,prb,showpacs,preprintnumbers,amsmath,amssymb,superscriptaddress]{revtex4}
\usepackage{graphicx}

\begin{document}
\title{Evidence of a magnetic Bose glass in \IPACBS~from neutron diffraction}
\author{Tao Hong}
\affiliation{Neutron Scattering Science Division, Oak Ridge National Laboratory, Oak Ridge, Tennessee 37831, USA}
\author{A.~Zheludev}
\email[Electronic address: ] {andrey.zheludev@psi.ch}
 \affiliation{Laboratorium für Festkörperphysik, ETH  Zürich, CH-8093
Switzerland}
 \affiliation{Laboratory for Neutron Scattering, ETH
Zürich and Paul Scherrer Institute, Villigen PSI, CH-5232
Switzerland}
\author{H.~Manaka}
\affiliation{Graduate School of Science and Engineering, Kagoshima
University, Korimoto, Kagoshima 890-0065, Japan}
\author{L.-P. Regnault}
\affiliation{CEA-Grenoble, INAC-SPSMS-MDN, 17 rue des Martyrs, 38054 Grenoble
Cedex 9, France}

\date{\today}

\begin{abstract}
We report the single crystal study of the bulk magnetization and
neutron scattering measurements on a quantum S=1/2 spin ladders
system \IPACBS~with quenched disorder. In zero field, the disordered
spin liquid phase is preserved as in pure \IPA. Due to the bond
randomness, a different Bose glass phase was directly observed in
$H_c$$<$$H$$<$$H'$, which separates the spin liquid phase from the
unconventional Bose-Einstein condensation phase. The observed finite
value of boson compressibility (\emph{dM/dH}) and lack of
field-induced three-dimensional long range order are consistent with
the theoretical prediction.
\end{abstract}

\vskip2pc

\maketitle

A ``Bose glass'' (BG) is an exotic state of matter that emerges in
systems of interacting bosons in the presence of quenched disorder.
At sufficiently low temperatures, {\it disorder-free} bosons are
subject to so-called Bose-Einstein condensation (BEC). BEC can
involve atoms in liquid $^4$He,\cite{London1938} laser-cooled ions
in magnetic traps,\cite{Wynar2000} Cooper pairs in superconductors,
\cite{BCS} or magnons in magnetic systems.\cite{Giamarchi2008} Due
to peculiarities of Bose statistics, particles lose their
individuality and occupy a unique quantum-mechanical state. The wave
function of this condensate establishes long-range quantum phase
coherence across a macroscopic sample. For repulsive bosons,
quenched disorder disrupts the condensate and interferes with phase
coherence. The result is a peculiar glassy state with only
short-range phase correlations.\cite{Giamarchi,Fischer1989} While
some experimental evidence of this was found in ultracold atoms,
\cite{Wang2004} high-temperature superconductors,\cite{Nelson1992}
 and quantum magnets,\cite{Oosawa2002-2,Manaka2008} none of the
 studies were direct. The
key characteristic, namely the wave function of the condensate
disrupted by disorder on the microscopic scale, remained
inaccessible. In this paper, we report a direct neutron diffraction
observation of short range correlations of the BEC order parameter
in a magnetic BG. This phase is realized in the quantum spin ladder
compound \IPACBS, where disorder is induced by random chemical
substitution.\cite{Manaka2001-2}

The disorder-free parent compound \IPA\ is a prototypical $S=1/2$ AF
spin ladder material with the ladders running along the
\textbf{\textbf{a}} axis of the crystal.\cite{Masuda2006,Zheludev2007}
 Nearest-neighbor spin interactions
along the legs of each ladder are AF. Nearest-neighbor inter-leg
correlations are ferromagnetic (FM). However, inter-leg coupling is
dominated by next-nearest-neighbor interactions. These are formed by
Cu-Cl-Cl-Cu superexchange pathways, and are robustly AF. As
discussed in Ref.~13, magnetic anisotropy in this
material is negligible. Zero-point quantum spin fluctuations in such
Heisenberg ladder structures destroy conventional magnetic order.
The result is a non-magnetic ``spin liquid'' state. The
lowest-energy excitations are a triplet of long-lived $S=1$
quasiparticles with a minimum excitation energy $\Delta$. For \IPA,
this energy gap is $\Delta=1.17$~meV.\cite{Masuda2006} The
quasiparticles obey Bose statistics, and are mutually repulsive at
short distances. Since the energy cost of creating each
quasiparticle is at least $\Delta$, the ground state is a vacuum of
bosons. The vacuum persists in modest applied magnetic fields.
However, due to Zeeman effect, the gap in the $S_z=+1$ magnon
decreases linearly with increasing field $H$, and reaches zero at
$H_c=\Delta/(g\mu_{\mathrm{B}})$. For \IPA\, $H_c=9.7$~T.
\cite{Manaka1998} Once $H>H_c$, the quasiparticle energy becomes
negative, and macroscopic number of them are incorporated in the
ground state. Since each carries a spin projection $S_z=+1$, their
density is equal to the uniform magnetization: $\langle \rho \rangle
=m\equiv\langle S_z \rangle$. {\it Simultaneously}, the emerging
bosons undergo magnon BEC.\cite{Giamarchi1999,Giamarchi2008} The
signature of this quantum phase transition is the appearance of
spontaneous long-range {\it staggered} (AF) magnetic order of spin
components {\it perpendicular} to the direction of applied field.
This transverse magnetization, written in complex form $\Psi=\langle
S_x \rangle + i \langle S_y \rangle$, is the effective wave function
of the Bose condensate. In  \IPA\ it was previously {\it directly}
probed by means of magnetic neutron diffraction, where the measured
scattering intensity is proportional to the Fourier transform of the
spin correlation function.\cite{Garlea2007,Zheludev2007} The BEC
phase is characterized by a new set of magnetic Bragg peaks with
half-integer Miller indexes. Their intensity is proportional to the
square of the BEC order parameter. Note that in other experimental
realizations of BEC of magnons, such as that in thin
films\cite{Demokritov} or $^3$He\cite{Bunkov}, the condensate wave
function remains experimentally inaccessible.

The best way to introduce quenched disorder in a magnetic system is
by chemical substitution. Most previous studies targeted the
magnetic ions, randomly substituting them by non-magnetic or
different spin impurities. The problem with this approach is that it
qualitatively alters the nature of the spin liquid state. Upon
substitution, local $S=1/2$ degrees of freedom are liberated in
direct proportion to the impurity concentration.\cite{Affleck1987}
These free spins are the dominant contribution to bulk
magnetization, give rise to a divergent magnetic susceptibility, and
enable conventional long-range ordering at low temperatures.\cite{Uchiyama1999}
In addition, each impurity becomes a potent
scattering center. This causes a collision damping of magnons,\cite{Regnault1995}
 making the very quasiparticles that are
supposed to condense poorly defined. Thus, any physics related to
the BEC transition is wiped out. With this in mind, in the present
study we adopted a radically different ``softer'' method. Quenched
disorder was introduced in \IPA\ by a partial substitution of
non-magnetic Br$^-$ for the likewise non-magnetic Cl$^-$.\cite{Manaka2008}
 This modification does not directly involve the
spin-carrying Cu$^{2+}$ ions and does not add additional anisotropy
term. Instead it affects the bond angles in the
Cu-halogen-halogen-Cu superexchange pathways. The strength of
magnetic interactions on the affected bonds is thereby modified, but
their AF character is preserved.

\IPACBS\ single crystals were grown in solution as described in
Ref.~11. The microscopic homogeniety of
Br-distribution was confirmed by single crystal X-ray diffraction
studies, that also confirmed the Br content to be within 10\% of
nominal. Bulk magnetization data were taken at \emph{T}=500~mK. The
magnetic field was applied along the $c$ axis of the crystal.\cite{Manaka2008}
 Neutron experiments were carried out in the
$(h,k,0)$ scattering plane on an assembly of five fully deuterated
crystals with a total mass of 0.8~g and a mosaic spread of
0.7$^\circ$. Inelastic measurements were performed on IN22 3-axis
spectrometer at ILL, using a standard He-flow cryostat with a Be
neutron filter after the sample and fixed final neutron energy
$E_f$=5~meV. High-field diffraction data were collected using a
vertical field cryomagnet and a dilution refrigerator insert on the
IN22, with a Pyrolytic Graphite (PG) filter after the sample and
$E_f$=14.7~meV. The experimental resolution was calculated in the
Popovici approximation.

\begin{figure}
\includegraphics[width=7.5cm,bbllx=10,bblly=20,bburx=575,
  bbury=370,angle=0,clip=]{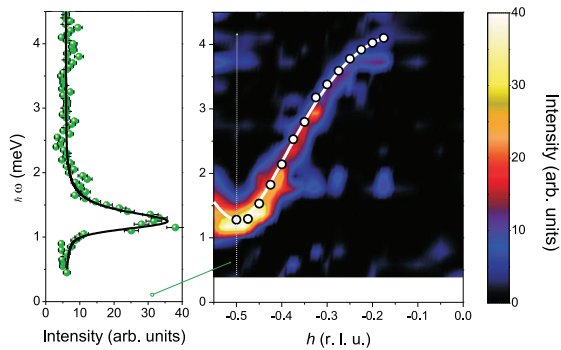}
\caption{(color online). The excitation spectrum (false color plot) measured in \IPACBS\ at $T=1.5$~K in zero applied field reveals sharp dispersive quasiparticles with an energy gap $\Delta=1.24(1)$~meV. A typical energy scan (left panel, symbols) shows  a well-defined peak with an energy width entirely due to experimental resolution (solid line, calculation).} \label{fig1}
\end{figure}

Previous bulk studies have shown that in \IPACB\ the spin liquid
ground state {\it remains intact} up to about $x=13\%$ Br-content.\cite{Manaka2001-2}
 In the present work we shall focus on the
$x=0.05$ material. While magnetic susceptibility experiments
revealed some residual free spins in this system, their
concentration is negligibly small.\cite{Manaka2001-2} From the bulk
magnetization, where the paramagnetic contribution at $T=500$~mK
remains smaller than 0.003~$\mu_\mathrm{B}$ per formula up to
$H=8$~T, we estimate that there are fewer than one free $S=1/2$ spin
for every 10 Br substitutions.\cite{Manaka2008} These are likely
due to crystallographic defects: \IPACBS\ crystals grown from
solution are systematically smaller and of inferior quality compared
to those of the pure compound. The singlet ground state in
IPA-Cu(Cl$_{0.95}$Br$_{0.05}$)$_3$ was confirmed in recent mu-SR
studies, where no long-range magnetic order was observed down to at
least $T=330$~mK.\cite{Saito2006} However, the clearest evidence of
that this material is an immaculate spin liquid is provided by our
recent inelastic neutron scattering experiments. The excitation
spectrum shown in Fig.~\ref{fig1} was measured at $T=1.5$~K and
reveals well-defined  bosonic quasiparticles with a spin gap
$\Delta=1.24(1)$~meV. Just like in the pure \IPA\cite{Masuda2006}
and PHCC\cite{Stone2006}, the quasiparticle spectrum terminates at
a critical wave vector $h_c\sim -0.2$. Apart from the slightly
larger gap energy, these excitations are almost indistinguishable
from those in pure \IPA. They are also at least as long-lived:
knowing our energy resolution, we can estimate the intrinsic
quasiparticle energy width to be $\Gamma<0.03$~meV. Thus, despite
the structural disorder, \IPACBS\ indeed {\it remains a true quantum
spin liquid}.

\begin{figure}
\includegraphics[width=7.5cm,bbllx=10,bblly=10,bburx=445,
  bbury=485,angle=0,clip=]{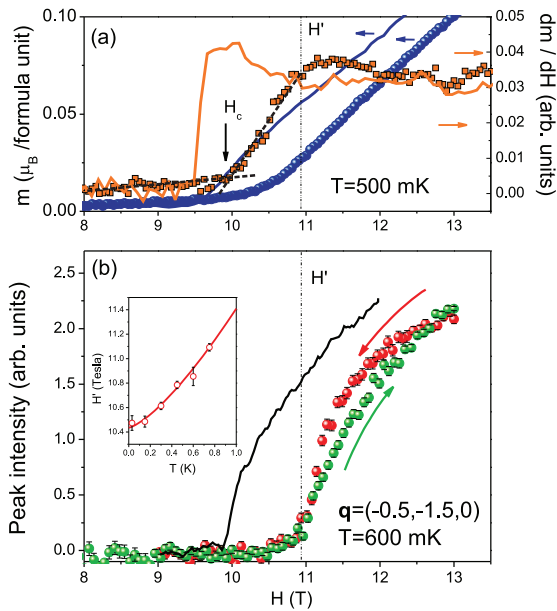}
\caption{(color online). Signatures of various quantum phases in \IPACBS.\ The
uniform magnetization (a, circles, from Ref. \protect\cite{Manaka2008}) represents the density of bosons in the ground state. The rate of increase of uniform magnetization is the boson compressibility (a, squares). The thin solid lines correspond to the disorder-free parent compound~\IPA. (b) The magnetic Bragg intensity measured at the maximum represents the effective BEC order parameter. The measured temperature dependence of $H'$ is shown in the inset. The solid curve is a guide for the eye.} \label{fig2}
\end{figure}

In external magnetic fields exceeding $H_c\sim 10$~T, \IPACBS\
becomes magnetized (Fig.~\ref{fig2}(a), circles).\cite{Manaka2008}
The magnetization derivative does not jump abruptly, as in \IPA\
under similar conditions (Fig.~\ref{fig2}(a), solid curves).
Instead, it gradually increases between $H_c$ and $H'\sim 11$~T
(Fig.~\ref{fig2}(a), squares), and is roughly constant at higher
fields. $H_c$ was determined by linear interpolations of the $dM/dH$
curve, as shown in Fig.~\ref{fig2}. $H'$ was determined in power-law
fits to the temperature dependence of the $(0.5, 1.5, 0)$ magnetic
Bragg peak measured upon cooling. The net magnetization and its
derivative are to be interpreted as the density and compressibility
of $S_z=+1$ quasiparticles now present in the ground state. Our key
result is that, unlike in the parent compound, in the doped material
these quasiparticles initially {\it fail to form a condensate}, the
latter only setting in at higher fields. In the pure system, sharp
Bragg reflections corresponding to magnetic long-range order appear
simultaneously with bulk magnetization (Fig.~\ref{fig2}(b), solid
curve).\cite{Garlea2007,Zheludev2007} In contrast, in \IPACBS\,
only broad peaks are observed around the propagation vector
$(1/2,1/2,0)$. At $T=600$~mK they are totally absent below $H_c$,
appear somewhere between $H_c$ and $H'$ but remain barely
detectable, and grow rapidly beyond $H'$ (Fig.~\ref{fig2}(b),
circles). They persist to the maximum attainable experimental field
of 13~T. Similar behavior is observed at other temperatures. The
measured $T$-dependence of $H'$ is plotted in the inset of Fig.~2.

\begin{figure}
\includegraphics[width=7.5cm,bbllx=5,bblly=10,bburx=410,
  bbury=570,angle=0,clip=]{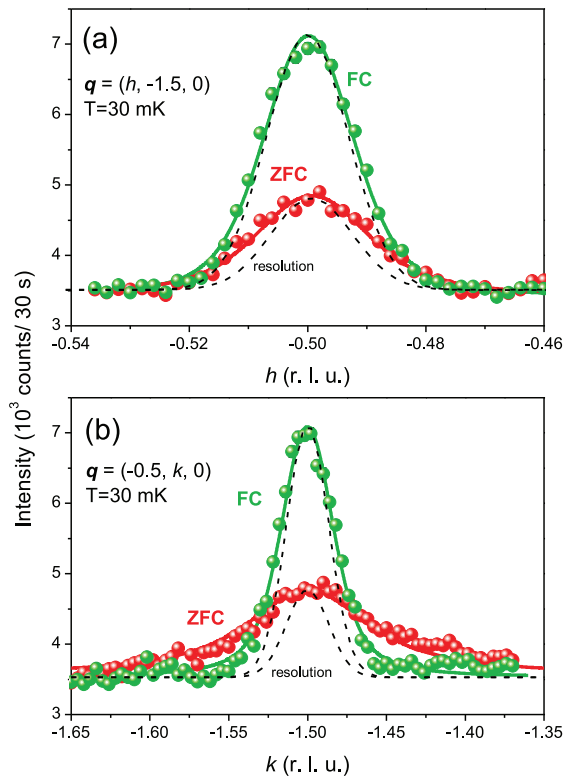}
\caption{(color online). Neutron scans across the $(-0.5, -1.5,0)$ point measured in zero-field-cooled (ZFC) and field-cooled (FC) \IPACBS\ samples at $T=30$~mK in $H=13$~T applied field (symbols), in the BEC phase. The magnetic peaks are broader than experimental resolution (dashed lines), indicating that only short-range correlations of the BEC order parameter are present. The solid lines are Voigt function fits to the data.} \label{fig3}
\end{figure}

The main panel of Fig.~\ref{fig3} shows scans across one
representative reflection. Here the dashed lines are the
experimental resolution. The finite intrinsic peak widths
unambiguously indicate that ordering is short-range. The
corresponding correlation lengths are history-dependent. This is
because in order to enter the BEC phase the system has to cross
through the disordered BG state. The resulting pinning of magnetic
domain walls or vortexes by the random potential makes true
long-range order kinetically inaccessible. In a sample
zero-field-cooled (ZFC) to 30~mK, at $H=13$~T, the correlation
lengths are $\zeta_a=70(6)$ lattice units along the ladder direction
($a$-axis of the crystal) and only $\zeta_b=11.7(0.1)$ lattice units
perpendicular to it ($b$-axis). If a sample is cooled in a 13~T
field, the peaks are much sharper, though still broader than
resolution, with $\zeta_a=143(8)$ and $\zeta_b=52(2)$. The peak
intensity is correspondingly higher in the field-cooled (FC) sample,
and shows hysteresis when the field is repeatedly decreased below
$H_c$, then again increased (Fig.~\ref{fig2}(b)). Similar hysteresis
was observed in \IPACBS\ at all temperatures between 30 and 750~mK,
but never in pure \IPA.\cite{Garlea2007,Zheludev2007}
History-dependent behavior and a difference between FC and ZFC
samples are a signature of a magnetic glass.

Based on bulk magnetization data, it was previously suggested that a
BG emerges in \IPACBS\ in magnetic fields exceeding $H_{c3}\sim
40$~T, while for $H_c$$<H<H_{c3}$ the system is in a BEC phase.\cite{Manaka2008}
 The present direct measurement of the effective
BEC wave function correlations unambiguously show that the glass
forms already at $H_c$, as soon as the bosons precipitate in the
ground state. This behavior can be qualitatively understood in the
framework of recent theories.\cite{Nohadini2005,Roscilde2006} The
ground state of weakly interacting spin ladders are a linear
combination of valence-bond states composed of local singlet spin
pairs. Due to the spatially randomized interaction strength, certain
singlets may have a reduced gap energy. When the magnetic field is
increased past $H_c$, these are the first to become magnetized. Each
such broken singlet corresponds to a $S_z=+1$ boson localized by the
quenched disorder. The phenomenon is similar to Anderson
localization of fermions in a random potential. The role of the
Pauli exclusion principle that enables the latter is played by the
strong short-range repulsion between spin ladder excitations. The
result is the magnetic BG with a finite boson density, but only
short-range frozen transverse spin (effective wave function)
correlations.\cite{Giamarchi,Fischer1989} A lack of long-range
magnetic order and a non-zero yet finite compressibility ($dM/dH$)
are the two key feature of a BG. The latter distinguish it from
other disordered phases of quantum spin systems: the incompressible
Mott glass{\cite{Giamarchi2001}} and the random single phase that
has divergent compressibility\cite{Altman2008}. The quasi-1D
character of \IPACBS~undoubtedly favors boson localization and helps
stabilize the BG state. Indeed, only in one dimension does the BG
phase appear for arbitrary weak disorder.\cite{Giamarchi} In higher
dimensions the magnitude of disorder needs to exceed some threshold
value to disrupt the condensate.\cite{Fischer1989}

At a certain higher field, that we identify with $H'$, the gaps
associated with even the strongest bonds in the system are overcome
by Zeeman energy. Beyond this point one expects to recover the
coherent BEC phase,\cite{Nohadini2005} in which the compressibility
is constant and AF spin correlations are long-range.\cite{Giamarchi2008}
 Qualitatively, the field range $H'-H$
corresponds to the energy difference between the strongest and
weakest AF bonds. The same energy scale determines the
doping-induced shift of the gap at $H=0$. For \IPACBS\ the latter is
of the order of 0.1 meV, and corresponds to a field range of 1~T,
consistent with the measured $H'-H_c$.

In our experiments, beyond $H'$,  the compressibility indeed levels
off and AF correlations rapidly build up. However, the latter remain
short-range. In this system, there is not necessarily a
contradiction. We suggest that while at higher fields the BEC phase
may be the true ground state, at low temperatures it remains
kinematically inaccessible. As the external field is increased
beyond $H_c$, local correlated regions grow in size around each
broken singlet that act as nucleation centers. By the time $H'$ is
reached, the macroscopic sample is a mosaic of uncorrelated finite
size AF domains or a textured pattern of magnetic vortexes. At low
temperatures the domain walls and/or vortexes are pinned by the
random potential. The system is frozen in this short-range
correlated state. In a field-cooled sample, the BG-BEC boundary is
crossed at a higher temperature, where the domain walls and vortexes
are more mobile. The result is fewer pinned defects, longer-range
correlations and and sharper diffraction peaks. This interpretation
allows us to reconcile the conclusions of previous bulk measurements\cite{Manaka2008}
with the present neutron scattering study. For
$0<H<H_c$ \IPACBS\ is a true spin liquid. For $H_c<H<H'$ the system
becomes a magnetic BG. Long-range order in the form of BEC may be
the ground state for $H'<H<H_{c3}$, but the system stays frozen in a
metastable glassy state. Between $H_{c3}$ and the saturation field
$H_2\sim 60$~T a second BG phase is realized, as discussed in
Ref.~10. It is related to the saturation transition,
that can also mapped on BEC.\cite{Batyev1984}

In summary, the unique ability of neutron scattering to probe the
effective condensate wave function correlations and of magnetization
measurements to probe boson density and compressibility in
magnetized quantum spin liquids, allowed us to directly observe the
exotic magnon BG phase.

\begin{acknowledgments}
One of the authors T.~Hong would thank G.~W.~Chern for the helpful
discussion. This work was supported (in part) under the auspices of
the United States Department of Energy. Research at ORNL was
supported by the DOE BES Division of Scientific User Facilities. One
of the authors H.~Manaka is supported by a Grant-in-Aid for Young
Scientists (B) from the Ministry of Education, Culture, Sports,
Science and Technology (MEXT).
\end{acknowledgments}


\end{document}